\documentclass[preprint,12pt]{elsarticle}

\usepackage{graphicx}
\usepackage{amsfonts}
\usepackage{amsmath, amsthm, amssymb}

\usepackage[utf8x]{inputenc}
\usepackage{geometry}

\newcommand{\R}{\mathbb{R}}

\newcommand{\erf}{\mathrm{erf}}
\def \E {\mathbb{E}}

\newcommand{\x}{\mathbf{x}}

\newcommand{\I}{\mathbf{I}}
\newcommand{\C}{\mathbf{C}}
\newcommand{\M}{\mathbf{M}}

\newcommand{\Id}{\mathbf{I}}

\def \M {\mathbf{M}}

\def \I {\mathcal{I}}
\def \L {\mathbf{L}}
\def \E {\mathbf{E}}
\def \V {\mathbf{V}}

\def \a {\mathbf{a}}

\def \R {\mathbb{R}}
\def \a {\mathbf{a}}
\def \C {\mathbf{C}}

\def \vv {\mathbf{v}}

\def \u {\mathbf{u}}
\def \x {\mathbf{x}}

\def \W {\mathbf{W}}
\def \C {\mathbf{C}}

\journal{Neural Networks}
\begin{document}
\begin{frontmatter}

\title{Context-dependent representation in recurrent neural networks}

\author{Gilles Wainrib}

\address{Ecole Normale Superieure, Departement d'Informatique, Paris, France.}

\begin{abstract}
In order to assess the short-term memory performance of non-linear random neural networks, we introduce a measure to quantify the dependence of a neural representation upon the past context. We study this measure both numerically and theoretically using the mean-field theory for random neural networks, showing the existence of an optimal level of synaptic weights heterogeneity. We further investigate the influence of the network topology, in particular the symmetry of reciprocal synaptic connections, on this measure of context dependence, revealing the importance of considering the interplay between non-linearities and connectivity structure.
\end{abstract}

\begin{keyword}
Sort-term memory, Echo-state networks, Recurrent neural networks, Random matrix theory, Mean-field theory, Information
\end{keyword}

\end{frontmatter}

\section{Introduction}

\begin{sloppypar}
The ability of the neural representation of a signal to depend on its past context and the capacity of short-term memory are essential for most perceptive and cognitive processes, from vision to langage processing and decision making \cite{jonides2008mind,todd2004capacity,boutla2004short,Bernacchiaetal11}. Although adaptation and plasticity may play a crucial role to shape short-term memory, we study here the hypothesis that this phenomenon may be also explained by dynamical network effects in particular due to the recurrent nature of neural networks connectivity \cite{compte2000synaptic,wang2001synaptic,major2004persistent,mante2013context}. To study this question from a theoretical standpoint, we consider the following classical model of recurrent neural network (RNN)\cite{wilson1972excitatory, amari1972characteristics,sompolinsky1988chaos,jaeger2004harnessing}, often called \textit{rate model}:\end{sloppypar}
\begin{equation*}\label{eq:main}
	\x(t+1) = S(\W\x(t) + \V\u(t) + \eta(t))
\end{equation*}
where $\x(t)\in \R^n$ describes the states of the $n$ neurons, $\u(t)\in \R^m$ represents the signal, $S(.)$ is typically a sigmoid function (\emph{e.g.} $\tanh(.)$), $\M$ is a $n \times m$ matrix projecting the input into the recurrent network, and $\W$ is a $n \times n$ matrix representing the internal connectivity of the recurrent network ($\W_{ij}$ is the connection strength from neuron $j$ to neuron $i$, and is often called \emph{synaptic weight}). We will make various assumptions on the matrices $\V$ and $\W$, and in particular we will investigate the disordered system where $\V_{ij}$ are independent $\mathcal{N}(0,\kappa^2 / m)$, and $\W_{ij}$ are either independent $\mathcal{N}(0,\sigma^2 / n)$ (random asymmetric connectivity) or constrained to be symmetric with $\W_{ij}=\W_{ji} \sim \mathcal{N}(0,\sigma^2/4n)$ (random symmetric connectivity). The parameter $\sigma$ characterizes the synaptic weights heterogeneity and is one of the most important parameter in this study because it controls the amount of recurrence in the generation of the representation. The variables $\eta(t)$ are independent random centered Gaussian vectors in $\R^n$ with diagonal covariance $\epsilon^2\mathbf{I}$. In this model, we have assumed a form of "network noise" that is associated with the network dynamics. However, it is also possible to consider at least two other natural sources of noise, namely an "input noise" in which $\u(t)+\eta(t)$ replaces $\u(t)$, or an "output noise" in which the observation of the network state $\x(t)$ is perturbed by a measurement noise $\x(t)+\eta(t)$. 

Most attempts to study theoretically short-term memory in similar random neural networks models have been focused on the linear case $S(x)=x$ because of the difficulty to handle this question in the non-linear regime. In the context of machine learning applications, \cite{jaeger2001} has shown the ability of this system to reconstruct faithfully up to $n$ time-steps in the past. In \cite{ganguli}, Fisher Information was used to assess how short-term memory depends upon the connectivity structure and the network size (see below). More recently, \cite{charles, sompolinsky} have studied the relevance of applying compressed sensing concepts to this problem, showing the possibility for very long short-term memory, scaling exponentially with $n$, for the recovery of sparse signals.

Before investigating the interplay between non-linearity and connectivity properties, which is the principal objective of the present article, we study statistical and information-theoretic measures (section 2.1) of the dependence of $\x(t)$ upon past stimulus $\u(t-k)$ for the linear model, and further introduce a new measure, the \textit{context capacity} (section 2.2), that will be easier to study in the non-linear case (section 3).

\section{Linear model $S(x)=x$}

\textbf{Preliminary results.}
For simplicity, we consider in this section the case of 1D input time-series, i.e. $m=1$ and $\V=\vv \in \R^n$. We introduce the transfer matrix: 
\begin{equation*}
\M=[\vv,\W\vv,\W^2\vv,...,\W^{T-1}\vv]
\end{equation*}
With this notation, in the absence of noise $\x(T)$ can be written as $\x(T) = \M \u$ where $\u$ is the $T$-dimensional vector $(u(0),...,u(T-1))$. Then, in the case of network noise 
$
\x(T) = \M \u + \sum_{k=0}^{T} \W^k \eta(t-k)
$
, in the case of input noise 
$
\x(T) = \M (\u+\eta) 
$
and finally in the case of output noise
$
\x(T) = \M \u+\eta
$. From these expressions, it is possible to estimate the covariance structure of $\x(T)$, assuming that the time-series $\u$ is a Gaussian process with zero mean and covariance matrix $\C$:
\begin{equation*}
	\E\left[\x(T)\x(T)'\right] = \M \C \M' + \Omega
\end{equation*}
where $\Omega$ is defined according to the type of noise considered:
\begin{itemize}
\item Network noise : $\Omega = \epsilon^2 \sum_{k=0}^{T-1} \W^k\W'^k$
\item Input noise : $\Omega = \epsilon^2 \M\M'$
\item Output noise : $\Omega = \epsilon^2 \mathbf{I}$.
\end{itemize}

\subsection{Some statistical and information-theoretic measures}
From the expression $\x(T)=\M\u$ in the absence of noise, one observes readily that when $T=n$, if the square matrix $\M$ is invertible (which is almost surely the case with random connectivity matrices), then the time-series $\u$ of length $T$ can be recovered exactly with the observation of $n=T$ neurons. However, this basic result does not take into account the impact of noise on the representation. In other words, if the matrix $\M$ is not well conditioned, then a small perturbation can lead to huge errors.

 The Cramer-Rao bound states that the variance of the reconstruction error for any estimator of $\u$ cannot be smaller than the inverse of the Fisher Information, therefore providing a universal bound for the input recovery problem. This measure of short-term memory has been studied in a remarkable work by \cite{ganguli} for the linear model with network noise, where it was shown that the Fisher matrix is given by
$$
\I_{kl} = \vv'\W'^{k} \Omega^{-1} \W^l \vv
$$
Here, the diagonal element $\I_{kk}$ is the Fisher information that $\x(t)$ contains about the input at time $t-k$, and characterizes the memory decay of the network representation. Interestingly, this measure of memory is independent of the input and characterizes only the recurrent neural network. After rescaling by the noise level $\I \equiv \epsilon^2 \I$, one defines the total memory
$
\bar{\I} = Tr(\I)
$ which satisfies the following fundamental distinction: if $\W$ is a normal matrix, then $\bar{I} = 1$, and otherwise, $\bar{I} \leq n$ and may behave extensively with network size. This result shows that the underlying structure of the connectivity $\W$ may have a profound impact on the dependence of the representation upon past context. In particular, random symmetric (normal) connectivity matrices  appear to be less efficient than asymmetric ones (non-normal) in terms of short-term memory. 

To further quantify this statement, one can evaluate the mutual information between $\x(T)$ and $\u$,
\begin{eqnarray*}
I(\x(T);\u) &=& \frac 1 2 \log \frac{\left| \det \left( \Omega + \M \C \M' \right)\right|}{|\det\Omega|}\\
&=&\frac 1 2 \log \left| \det \left(\Id_n + (\Omega^{-\frac 1 2}\M) \C (\Omega^{-\frac 1 2}\M)' \right)\right|\\
&=&\frac 1 2 \log \left| \det \left(\Id_T + (\M\C^{\frac 1 2})' \Omega^{-1} (\M\C^{\frac 1 2}) \right)\right|
\end{eqnarray*}
which is related to the Fisher information by\footnote{notice that a similar formula can be found in \cite{ganguli}.}:
$$
I(\x(T);\u) = \frac 1 2 \log | \det (\Id_T + \C^{\frac 1 2} \I \C^{\frac 1 2})|
$$ In FIG. 2 (left), the mutual information is displayed as a function of $\sigma$ for both the symmetric and asymmetric model, showing a clear superiority of the asymmetric model. To understand theoretically this observation, it is possible to use random matrix theory to evaluate the mutual information for $n\to \infty$, assuming $\C=\mu^2\Id_T$ and an output noise setting, for which $\Omega=\epsilon^2\Id$. When $\W$ is random asymmetric then
$$
I_{asym}(\x(T);\u) \sim \tilde{I}_{asym}=\frac 1 2 \sum_{k=0}^T \log \left(1+\frac{\mu^2}{\epsilon^2}n\sigma^{2k}\right)
$$ while the case of symmetric connectivity yields
$$
I_{sym}(\x(T);\u) \sim \tilde{I}_{sym}=\frac 1 2 \log \left| \det \left(\Id_T + \frac{\mu^2}{\epsilon^2} n\L\right)\right|
$$
where $\L$ is a checkerboard matrix filled with rescaled Catalan numbers:\\
$\L_{ij}  = C_{p-1} \left(\frac{\sigma}{2}\right)^{2(p-1)} \mbox{ if } i+j=2p \mbox{ with }C_p= \frac{1}{p+1}\binom {2p} {p}, $ and $\L_{ij}  =  0 \mbox{ if } i+j \mbox{ is odd.}
$\\
Indeed, to prove this result one only needs to evaluate the entries of the matrix $\M'\M$. For random asymmetric matrices, in the large $n$ limit:
$$\frac 1 n (\M'\M)_{ij} = \frac 1 n \vv'\W'^{i}\W^{j}\vv \to \delta_{ij}\sigma^{i+j}$$
whereas for random symmetric matrices:
$$\frac 1 n (\M'\M)_{ij} = \frac 1 n \vv'\W^{i+j}\vv \to \L_{ij}$$
where the Catalan numbers arise as the even moments of the semi-circular law.

The above determinant appears to be rather difficult to compute analytically, however it is still possible to compare the mutual information for these two cases. Applying Hadamard inequality to the determinant in $\tilde{I}_{sym}$:

$$
	\tilde{I}_{sym} \leq \frac 1 2 \sum_{i=1}^T \log \left( 1+ \frac{\mu^2}{\epsilon^2} nC_{i-1}\left(\frac{\sigma}{2}\right)^{2(i-1)}\right)
$$
and using the inequality $C_p \leq 4^p$, one concludes that the mutual information for symmetric linear RNN is smaller than the one for asymmetric case:
$$
	\tilde{I}_{sym} \leq \tilde{I}_{asym}.
$$
In fact, for any symmetric connectivity matrix with an eigenvalue distribution compactly supported in $[-\sigma,\sigma]$ (not only the semi-circular law), the same conclusion remains valid. Therefore, we have shown in this section that, in the linear model, RNN with symmetric connectivities capture less information about the past inputs than asymmetric ones.
\subsection{Context capacity}
All the previous estimations were heavily relying on the linear relationship between $\x$ and $\u$, enabling the use of linear algebra and (random) matrix tools. However, as noticed for instance in \cite{ganguli}, extending the above analysis to non-linear models is very challenging. To circumvent this difficulty, we introduce a new measure that quantifies how much the representation of a signal depends upon its past context, and that is amenable to analysis in the non-linear setting. To define our measure of context-dependence, we decompose the input time-series into two parts: the \emph{context} is the input from time $t=1$ to $t=t_0$, and the \emph{signal} is the input from time $t=t_0+1$ up to some time $t=t_0+\tau$. From this decomposition, we first define the \emph{context sensitivity} $\chi$ as follows (FIG. 1, left panel):\\ 
(a) we consider the context to be randomly generated according to a given probability law. At each trial, an independent source of noise $\eta$ is also generated.\\
(b) the signal is kept fixed to a specific time-series $\bar{\u}$\\
(c) we estimate the across-trial variance $\chi(\tau)$ of the representations $\x(t_0+\tau)$ obtained for each realization.\\
This variance can be explained by two sources of variability, namely the various contexts generated before the signal and the presence of noise in the construction of the representation. In order to normalize this variance with respect to the pure impact of noise, we introduce a measure of the variability of the representation due to noise only, called the \emph{unreliability} coefficient $\rho$ (FIG. 1, right panel):\\
(a) we consider the context to be a fixed time-series (say a fixed random sample generated according to the same given probability law), whereas t each trial, an independent source of noise $\eta$ is generated.\\
(b) the signal is kept fixed to a specific time-series $\bar{\u}$\\
(c) we estimate the across-trial variance $\rho(\tau)$ of the representations $\x(t_0+\tau)$ obtained for each realization.\\
Finally, we are in position to define the \emph{context capacity} $C(\tau)$ as the ratio of the context sensitivity over the unreliability coefficient:
\begin{equation*}
	\label{eq:def}
	C(\tau) = \frac{ \chi(\tau)}{ \rho(\tau)}.
\end{equation*}
If the representation is context-independent, then $C(\tau)=1$, and  the higher is $C(\tau)$, the more context-dependent is the representation, and moreover, one expects $C(\tau)$ to be a decreasing function of $\tau$.
 \begin{figure*}[t]
 	\begin{center}
		\includegraphics[width=8.7cm]{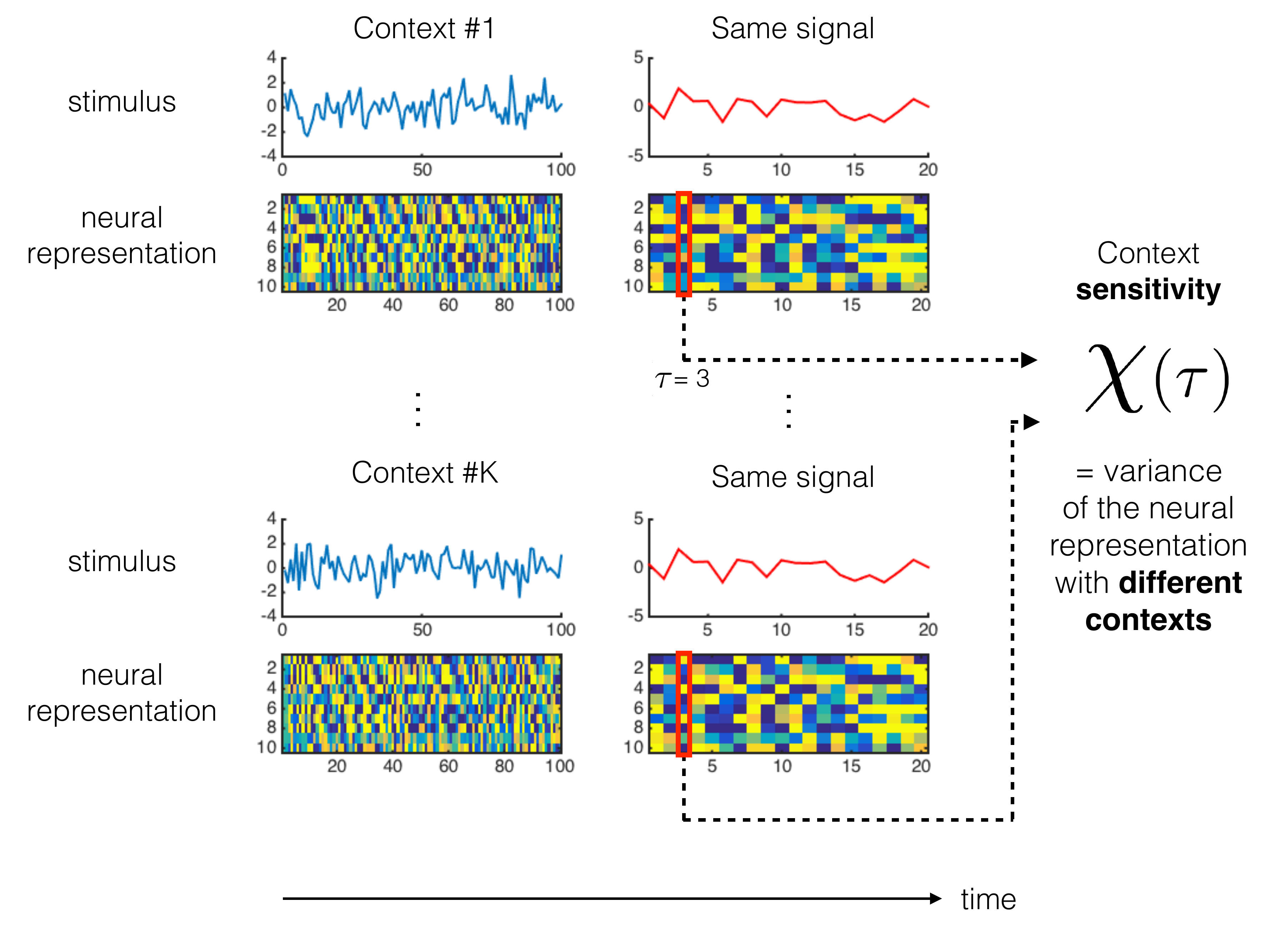}
		\includegraphics[width=8.7cm]{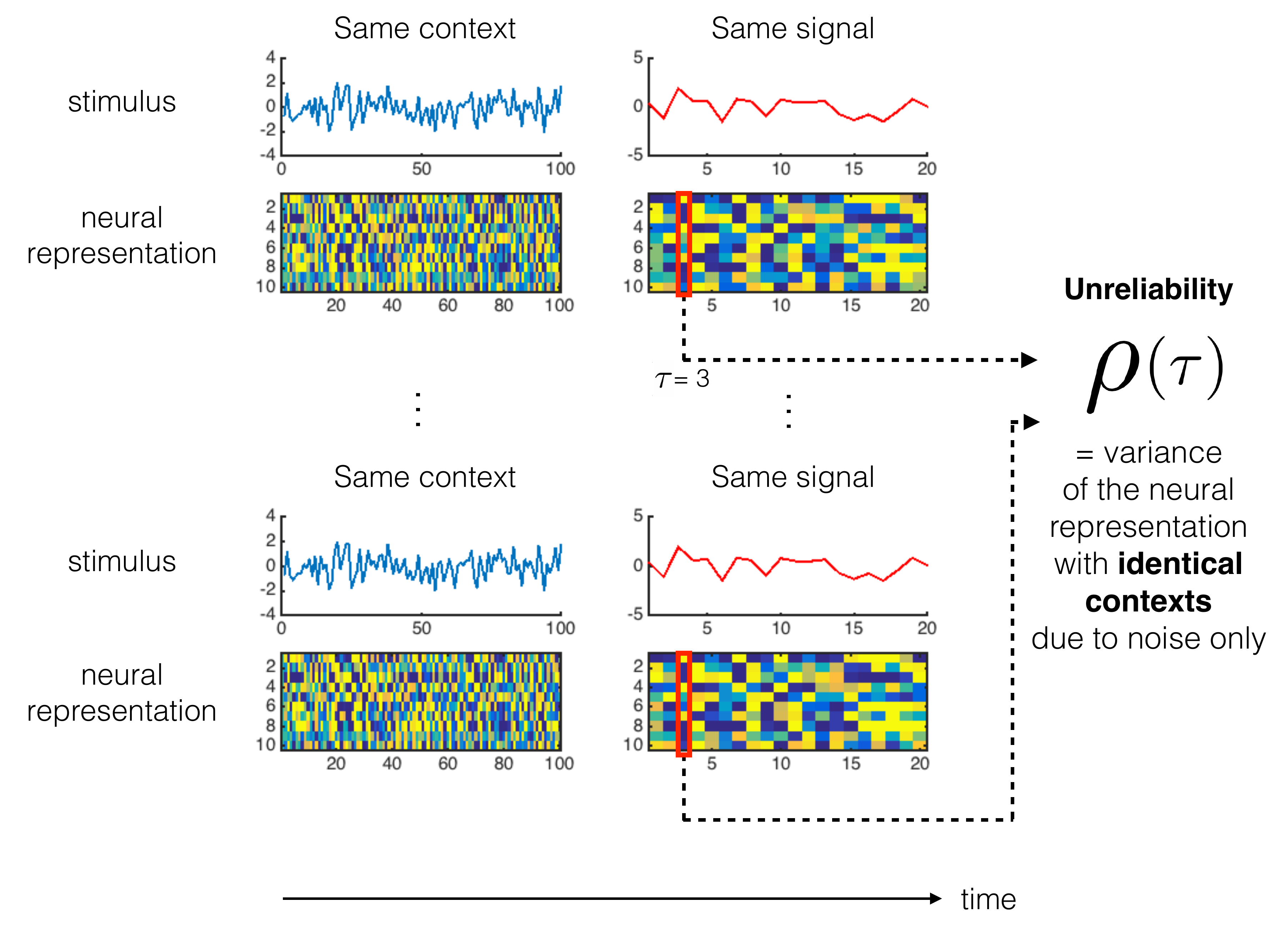}
 		\caption{Schematic illustration of the definition of the unreliability coefficient $\rho$ and context sensitivity $\chi$. }
 	\end{center}
	\label{fig:schematic}
 \end{figure*}
Before studying the non-linear system, we first evaluate the context capacity in the linear case to study its relationship with the above statistical and information-theoretic measures.

We make the assumption that $t_0$ is large. To compute the context sensitivity, we consider that, before time $t=t_0$, both the noise and the input are treated as random processes. Therefore $\x(t_0)$ is centered and has covariance 
$
	Cov(\x(t_0)) = \M\C\M' + \Omega
$. 
Then, at time $T=t_0 + \tau$, $\x(T)$ is no longer centered, and has covariance
$
	Cov(\x(T)) = \W^{\tau}\left(\M\C\M' + \Omega\right)\W'^{\tau} + \Omega_{\tau}
$ where $\Omega_{\tau}=\sum_{k=0}^{\tau-1} \W^k\W'^k$. Since $t_0 \to \infty$, one has the identity $\W^{\tau} \Omega \W'^{\tau} + \Omega_{\tau}= \Omega$, and the context sensitivity is given by  
$$
	\chi = Tr\left(\W^{\tau}\M\C\M' \W'^{\tau}+ \Omega\right)
$$ 
To compute the unreliability coefficient, the input is always considered as deterministic, so the covariance of $\x(T)$ is 
$
	Cov(\x(T)) = \Omega
$. 
Therefore, the unreliability coefficient is 
$$
	\rho = Tr\left(\Omega\right)
$$
Finally, the context capacity is given by:
\begin{equation*}
	C(\tau) =  1+ \frac{Tr\left(\W^{\tau} \M\C\M' \W^{'\tau} \right)}{Tr\left(\Omega\right)}
\end{equation*}
To investigate the impact of connectivity symmetry, we wish to analyze this formula from the RMT point of view, assuming $\C=\mu^2\Id$. First we need the following \textit{trace lemma}: in the limit $n\to \infty$, \begin{itemize}
\item for the asymmetric random model,
$
\frac 1 n Tr(\W^{k}\W'^{k}) \to \sigma^{2k}
$, 
\item while for the symmetric random model,
$
\frac 1 n Tr(\W^{k}\W'^{k}) \to C_{k}\left(\frac{\sigma}{2}\right)^{2k}
$.
\end{itemize}
Since $t_0\to \infty$, one first remark that:
$
	\W^{\tau}\M_{t_0}\M_{t_0}'\W'^{\tau} = \M_T\M_T' - \M_{\tau}\M'_{\tau}
$
where $\M_{k}$ is the extraction of the first $k\ $columns of $\M$.
Considering first the case of asymmetric connectivity, since $Tr(\M\M')=Tr(\M'\M)$ and since the diagonal terms of $\frac{1}{n}\M'\M$ converge to $\sigma^{2(i-1)}\mu^2$, then:
$$
Tr\left(\W^{\tau} \M\C\M' \W^{'\tau} \right) \sim n \mu^2 \left(\frac{1}{1-\sigma^2} - \frac{1-\sigma^{2\tau}}{1-\sigma^2}\right) = n\mu^2\frac{\sigma^{2\tau}}{1-\sigma^2}
$$
It remains to evaluate $Tr(\Omega)$, which can be done using the trace lemma, yielding:
$$
Tr(\Omega) \sim n \frac{1}{1-\sigma^2}
$$
Finally we obtain that when $n\to \infty$ and $T$ is fixed, 
\begin{equation*}
	\lim_{n\to \infty} C_{asym}(\tau)  =  1+ \frac{\mu^2}{\epsilon^2} \sigma^{2\tau} = \tilde{C}_{asym}(\tau) 
\end{equation*}
This formula is very simple and shows that the context capacity for the linear random asymmetric RNN is given by the product of the signal/noise ratio $\frac{\mu^2}{\epsilon^2}$ times a geometrically decaying term $\sigma^{2\tau}$ which is maximal when $\sigma$ approaches ones. Notice that the same product was already appearing in the expression of the mutual information $\tilde{I}_{asym}$.

The situation is, again, different for symmetric connectivity. Indeed, the diagonal terms of $\frac 1 n\M'\M$ now converge to $C_{i-1}\left(\frac{\sigma}{2}\right)^{2(i-1)}$, and we apply the trace lemma to obtain that when $n\to \infty$ and $T$ is fixed, 
\begin{equation*}
	\lim_{n\to \infty} C_{sym}(\tau)  =  1+\frac{\mu^2}{\epsilon^2} \frac{\Theta_{\tau}}{\Theta_0}= \tilde{C}_{sym}(\tau) 
\end{equation*}
where, for $\sigma<1$:
$$
	\Theta_{\tau} = \int_{-\sigma}^{\sigma} \frac{x^{2\tau}}{1-x^2} p(x)dx
$$
with $p(x)$ the density of the real eigenvalue distribution of $\W$. In the case of the semi-circular law, using the generating function of the Catalan numbers, one obtains:
$$\Theta_{0}	= \frac{2}{1+\sqrt{1-\sigma^2}}
\mbox{ and }
	\Theta_{\tau} = \Theta_0 - \sum_{k=0}^{\tau} \frac{C_k}{4^k}\sigma^{2k}$$
As for the mutual information $\tilde{I}_{sum}$, this formula involves Catalan numbers and is more explicit because the evaluation of the trace is more straightforward than the evaluation of the determinant.

From this result, using $C_k \leq 4^k$, we deduce that the context capacity for symmetric random model is lower than its asymmetric counterpart 
$
	\tilde{C}_{sym}(\tau) \leq \tilde{C}_{asym}(\tau)
$.
Therefore the context capacity behaves similarly as more standard measures such as the mutual information, and, as we will see in the next section, offers an interesting alternative to investigate the interplay between non-linearity and connectivity properties.

\section{Non-linear model}
\begin{sloppypar}
According to the mean-field theory \cite{sompolinsky1988chaos,cessac94,touboul,cabana2013large}, in particular in the case of input-driven systems \cite{rajan2010stimulus,massar2013mean,galtier2014local}, $\sigma$ is the most important parameter in this system since it controls an order-disorder phase transition between a "stimulus-driven regime" for $\sigma<\sigma_{critical}$ and a "chaotic regime" for  $\sigma>\sigma_{critical}$. It has been argued that the regime close to criticality in such systems may be relevant in terms of information processing capabilities, both in the fields of neuroscience \cite{beggs2008criticality} and  artificial intelligence \cite{boedecker2012information,bertschinger2004real}. Therefore, our first aim is to understand how this heterogeneity parameter affects the context capacity in the disordered neural network model. To measure the capacity $C$ we need to specify how we generate various contexts and select a given signal. For simplicity, we  assume that the input $\u(t)$ is a one-dimensional white noise process. In FIG. 2.(middle), we display the context capacity $C(\tau)$ for different values of $\tau$ as a function of the synaptic heterogeneity $\sigma$ for the random asymmetric model. As expected, the context capacity $C(\tau)$ is a decreasing function of $\tau$. More interestingly, it displays a maximal value for an intermediate value of $\sigma>1$, revealing a trade-off between recurrence-induced memory and non-linear instabilities. To understand this observation from a theoretical standpoint, our strategy is to develop a mean-field approximation of $C(\tau)$. To estimate the context sensitivity, we consider for each trial $k\in \{1,...,K\}$ the solution $\x^{(k)}$ of 
\end{sloppypar}
$$
	\x^{(k)}(t+1) = S(\W\x^{(k)}(t) + \V\u^{(k)}(t) + \eta^{(k)}(t))
$$
with the same initial condition $\x(0)$ and where $\eta^{(k)}$ are independent realization of the random process $\eta(t)$ defined above, and:\\
- for $1\leq t\leq t_0$, $\u^{(k)}(t)$ are independent standard Gaussian variables, representing various contexts,\\
- for $t_0+1\leq t\leq t_0+\tau$, for all trials, all the $\u^{(k)}(t)$ are equal to $\bar{\u}(t)$, which is a frozen realization of a white noise process.\\
In this problem, there are two different sources of randomness: the weights matrices $\W$ and 
$\V$ are randomly drawn once and for all, while the stochastic process $\eta$ and the various contexts are drawn at each trial $k$. The first source of randomness, although frozen, will be responsible, in the large $n$ limit, for a phenomenon of self-averaging, which is a well-established property of the mean-field theory \cite{sompolinsky,moynot,cessac}. More precisely, population averages of the form $\frac 1 n \sum_{i=1}^n f(\x_i(t))$ converge to the expectation $\E_{\W,\V}[f(\x_i(t))]$ over the law of the pair $(\W,\V)$, which is actually a quantity independent of $i$. 
 \begin{figure*}[t]

 	\begin{center}
		\includegraphics[width=5.5cm]{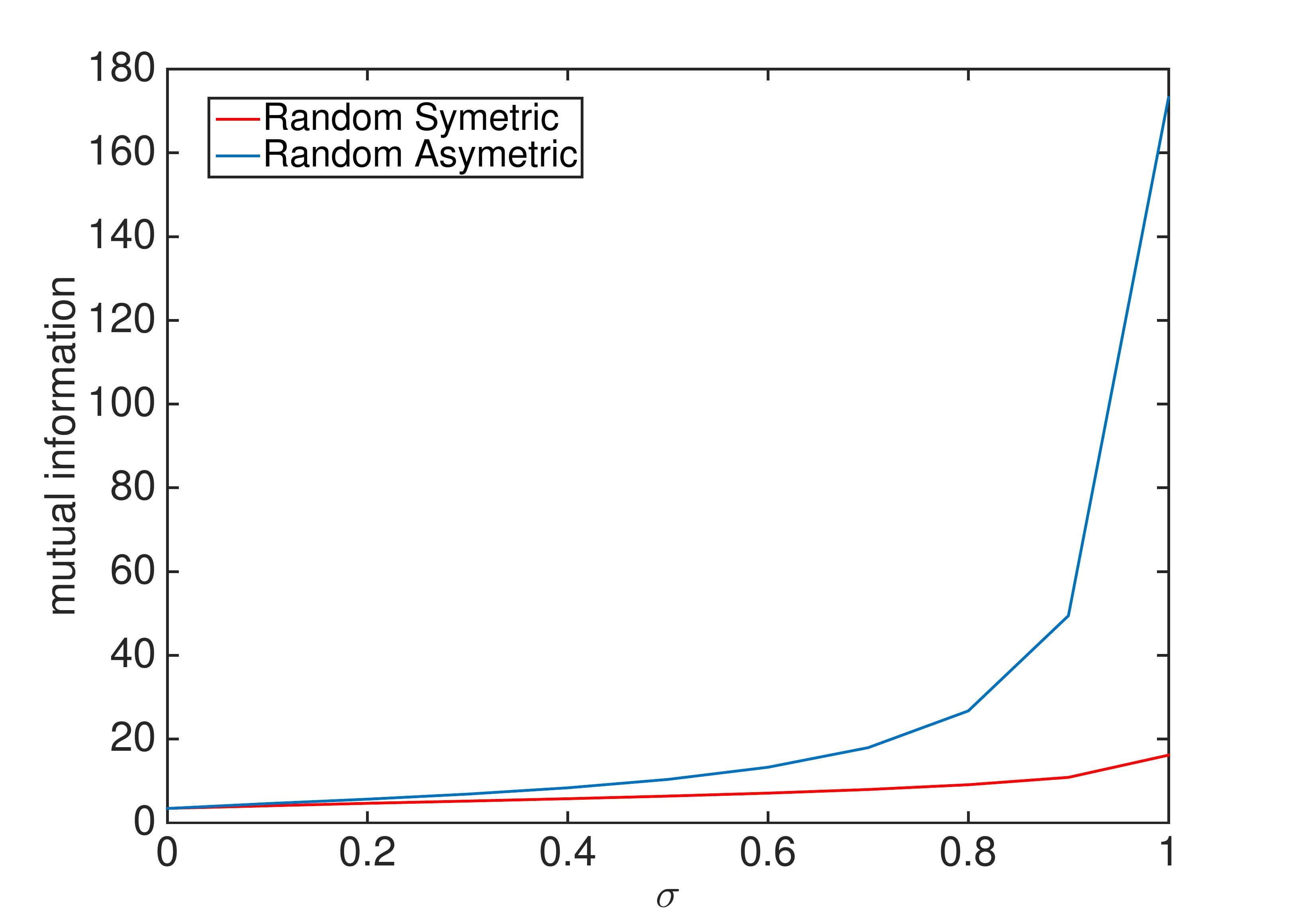}
 		\includegraphics[width=5.5cm]{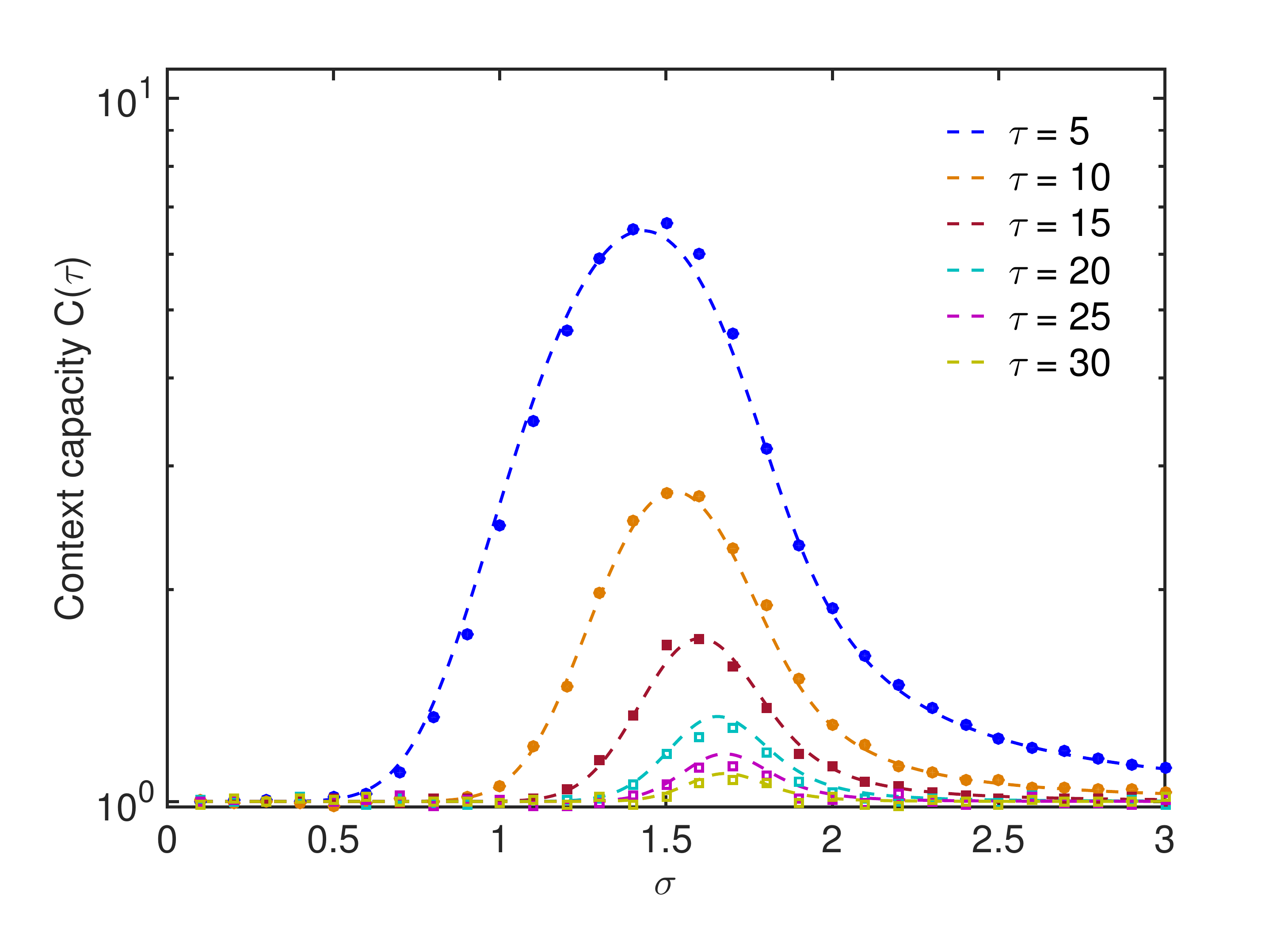}	
		\includegraphics[width=5.5cm]{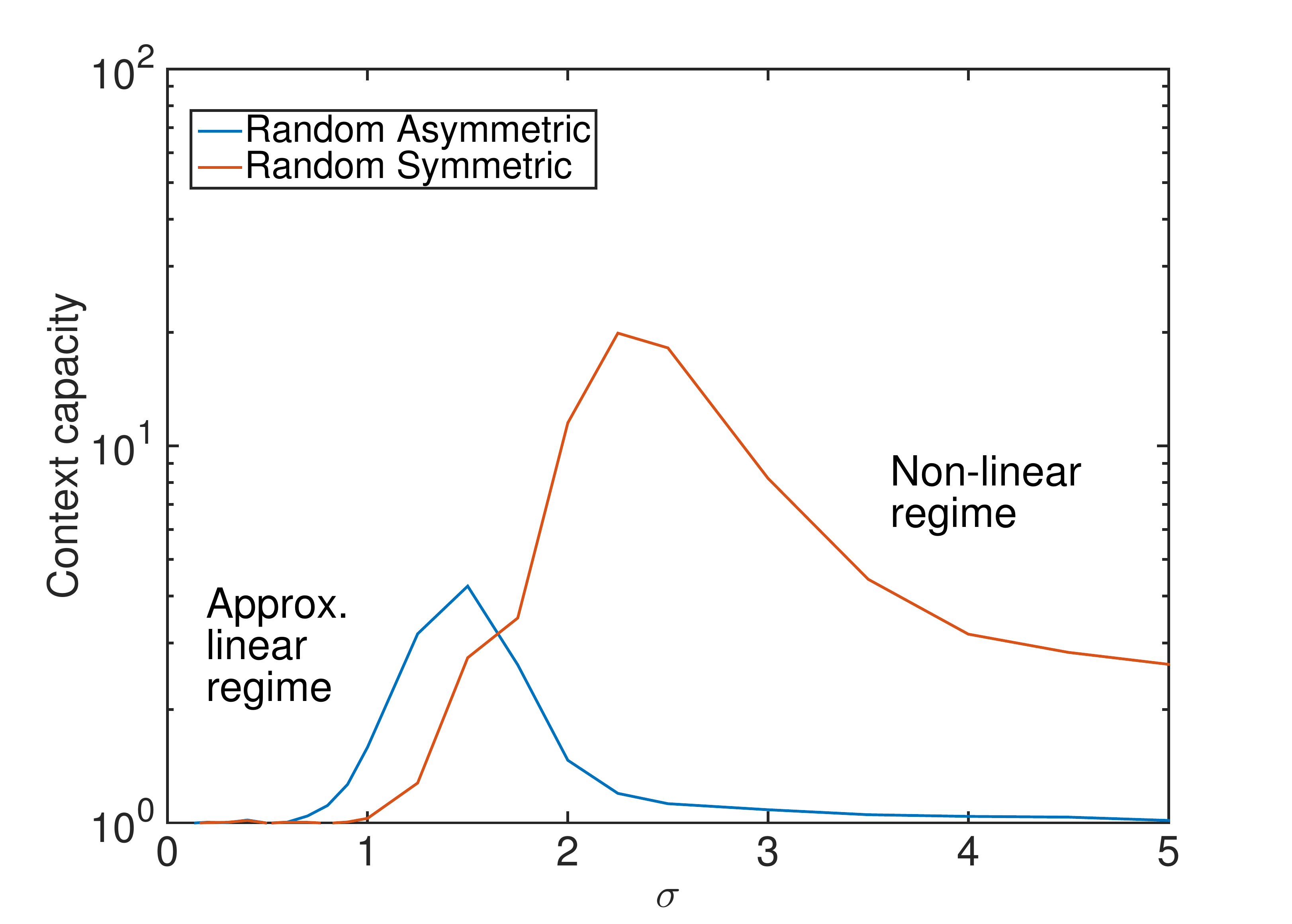}	
 		\caption{(\textbf{Left}) Mutual information as a function of $\sigma$ for the linear model (\textbf{Middle}) Context dependence capacity $C(\tau)$ as defined in \eqref{eq:def}, for the non-linear system with random asymmetric connectivity, as a function of the standard deviation $\sigma$ of the synaptic weights, for different values of the delay $\tau$. Points correspond to stochastic simulations with $n=1000$ neurons, and dotted lines correspond to theoretical predictions. (\textbf{Right}) Comparaison of the context dependence capacity $C(\tau=5)$ with respect to the symmetry of the connectivity. All numerical simulations of the non-linear model were done with $S(x)=\erf(\sqrt{\pi/2}x)$.}
 	\end{center}
	\label{fig:results}
 \end{figure*}
\\To estimate the across-trials variance of $\x_i^{(k)}(t)$, we need to compute:
$$
	v_i(t) := \langle\x_i(t)^2\rangle_K - \langle\x_i(t)\rangle_K^2
$$
where $\langle z\rangle_K= \frac 1 K \sum_{k=1}^K z^{(k)}$ denotes the across-trials average. Then we take the average over the neural population to obtain a scalar value $v(t)=[v]_N$ where $[z]_N=\frac{1}{n} \sum_{i=1}^n z_i$ denotes the population average. Introducing the sample covariance between two trials:
$
	\rho_{k,l}(t) := [\x_i^{(k)}(t) \x_i^{(l)}(t)]_N
$
we can rewrite:
$$
	v(t) = \langle\rho_{k,k}(t)\rangle_K - \langle\langle \rho_{k,l}(t)\rangle\rangle_K
$$
First, the trace term $\gamma(t):=\langle\rho_{k,k}(t)\rangle_K$ can be obtained using classical mean-field theory. Indeed, 
$$
	\gamma(t) = \langle[S(\a(t))]_N\rangle_K
$$
where the variables $$\a^{(k)}_i(t)=(\W\x^{(k)}(t) + \V\u(t) + \eta^{(k)}(t))_i$$ are asymptotically (for large $n$) independent Gaussian random variable with common zero mean and co-variance:
\begin{eqnarray*}
&(*) &	\E\left(\a^{(k)}_i(t^+)\a^{(l)}_i(t^+)\right) = \E\left( \sum_{a,b}\W_{ia}\W_{ib} \x_a^{(k)}(t)\x_b^{(l)}(t) \right)\\
	&+& \E\left( \sum_{a,b}\V_{ia}\V_{ib} \u_a^{(k)}(t)\u_b^{(l)}(t) \right)
	+ \E \left(\sum_{a,b}\eta^{(k)}_{a}(t)\eta^{(l)}_{b}(t)\right)
\end{eqnarray*}
where $t^+=t+1$. Here, we denote by $\E(.)$ the expectation with respect to the joint law of $(\W,\M,\eta,[\u]_1^{t_0})$. In fact, we only need to consider here the variance $\E\left(\a^{(k)}_i(t)\a^{(k)}_i(t)\right)$, which neither depend on $i$, since we took the average over $(\W,\M)$, nor on $k$ since we took the average over $\eta$ and $[\u]_1^{t_0}$. When considering $k=l$, the third term of the above sum is equal to $\epsilon^2$ by assumption on $\eta$, and the second one is given by $\kappa^2 u^2$. The first term is more problematic, since $\W$ and $\x$ could be, in principle, correlated. However, and this is the key point in the mean-field theory, this dependence decays when $n$ becomes very large, and in this asymptotic regime one can pretend $\x$ and $\W$ are independent (see \cite{moynot,cessac} for a rigorous justification, and \cite{massar2013mean} for a recent exposition of the application of the theory in the case of input-driven systems). Therefore, the first term in the sum can be approximated by $\sigma^2 \gamma(t)$, and we obtain formally:
$
	\E\left(\a_i(t)^2\right) = \sigma^2 \gamma(t) + \kappa^2 u^2 + \epsilon^2
$. 
Knowing the mean and variance of $\a^{(k)}_i$, since $\x(t+1)=S(\a(t))$, one can write a recurrent equation, also called the mean-field equation:
$$
	\gamma(t+1) = F(\sigma^2 \gamma(t) + u(t)^2 + \epsilon^2)
$$
where
$$
	F(x^2):= \frac{1}{\sqrt{2\pi}}\int_{\R} S(zx)^2 e^{-z^2/2} dz.
$$ 
Then, the second term to compute is the average of the sample covariance $\rho_{k,l}$ that we denote by $\lambda(t)$. The situation is slightly different: since
$
	\lambda(t+1) = \langle\langle[ S(\a_i^{(k)})S(\a_i^{(l)}) ]_N \rangle\rangle_K
$,
it appears here that knowing the variance of $\a_i^{(k)}$ is not sufficient, and we further need to estimate the covariance between $\a_i^{(k)}$ and $\a_i^{(l)}$ for $k\neq l$. The first term in the  sum $(*)$ is again given by $\sigma^2 \lambda(t)$, the second and third terms are equal to zero since all the trials are independent. Therefore, we obtain:
\begin{equation*}
	\lambda(t+1) = G(\sigma^2 \lambda(t) , \sigma^2 \gamma(t) + \kappa^2 u^2+ \epsilon^2)
\end{equation*}
where:
\begin{eqnarray*}
	G(x,y) &:=& \int_{\R^2} S(a_1)S(a_2)e^{-a'\Sigma^{-1}(x,y) a}da\\
	\Sigma(x,y) &:=&  \left( \begin{array}{cc}
	y & x \\
	x & y \end{array} \right)
\end{eqnarray*}
We have obtained a deterministic dynamical system describing the variance across trials $v(t)=\gamma(t) - \lambda(t)$, holding from time $t=1$ up to time $t=t_0-1$:
\begin{equation*}
  (E)\left\{
      \begin{aligned}
	\gamma(t+1) &= F(\sigma^2 \gamma(t) + \kappa^2 u^2 + \epsilon^2)\\
	\lambda(t+1) &= G(\sigma^2 \lambda(t), \sigma^2 \gamma(t) + \kappa^2 u^2 + \epsilon^2)
\end{aligned}
    \right.
\end{equation*}
To compute the context sensitivity coefficient $\chi$, one needs to solve the above dynamical system from time $1$ to $t_0-1$. Then, at time $t=t_0$, one switches to a system where $\u^{(k)}=\bar{\u}$ is now the signal and is the same for all trials, so that from time $t=t_0$ to $t=t_0+\tau$:
\begin{equation*}
  (E')\left\{
      \begin{aligned}
	\gamma(t+1) &= F(\sigma^2 \gamma(t) + \kappa^2 \bar{u}(t)^2 + \epsilon^2)\\
	\lambda(t+1) &= G(\sigma^2 \lambda(t) + \kappa^2 \bar{u}(t)^2, \sigma^2 \gamma(t) + \kappa^2 \bar{u}(t)^2 + \epsilon^2)
\end{aligned}
    \right.
\end{equation*}
To summarize, after solving (E), followed by (E'), one obtains $\chi(\tau) = \gamma(t_0+\tau)-\lambda(t_0+\tau)$.

To compute the unreliability coefficient $\rho$, the only difference is that the context $\u$ is now frozen and does not change across trials. For each trial $k\in \{1,...,K\}$, we consider $\x^{(k)}$ the solution of 
$$
	\x^{(k)}(t+1) = S(\W\x^{(k)}(t) + \M\u(t) + \eta^{(k)}(t))
$$
Therefore, the above derivation remains valid, with only minor modifications, yielding a slightly modified version of the mean-field dynamical system, for $t=1$ to $t=t_0-1$:
\begin{equation*}
  (E'')\left\{
      \begin{aligned}
	\gamma(t+1) &= F(\sigma^2 \gamma(t) + \kappa^2 u(t)^2 + \epsilon^2)\\
	\lambda(t+1) &= G(\sigma^2 \lambda(t)+\kappa^2 u(t)^2, \sigma^2 \gamma(t) + \kappa^2 u(t)^2 + \epsilon^2)
\end{aligned}
    \right.
\end{equation*}
and for $t=t_0$ to $t=t_0+\tau$ we obtain exactly the same system (E') defined above. Therefore, after solving (E''), followed by (E'), one obtains $\rho(\tau) = \gamma(t_0+\tau)-\lambda(t_0+\tau)$ and finally $C(\tau)=\chi(\tau)/\rho(\tau)$. In FIG. 2 (middle), theoretical predictions are compared with numerical simulations showing a good agreement.
\begin{sloppypar}
In light of the results obtained for the linear model, the next natural question is to compare the context capacity according to the connectivity properties, and in particular the symmetry of $\W$. So far, our theoretical results were based on mean-field theory, which heavily relies on the assumption of independent coefficients $\W_{ij}$ (asymmetric random model). Indeed, symmetry introduces a large amount of dependence in the matrix $\W$ and mean-field theory fails for the symmetric random model \cite{faugeras_mclaurin,faugeras20141,faugeras20142}. However, it is possible to evaluate numerically the context capacity in the symmetric case: as shown in FIG. 2 (right), $C$ is much much lower in the symmetric case for small values of $\sigma$, corresponding to an "almost-linear" regime, in accordance with results for the linear model, whereas it becomes much higher for larger values of $\sigma$, a regime where the non-linear effects become prominent. Indeed, in this regime, the asymmetric model displays chaotic dynamics hence a poor context sensitivity due to a high unreliability, while the symmetric model has an energy function \cite{hopfield}, which prevents chaos and ensures a higher context capacity. With this new observation, it appears that the subtle interplay between connectivity properties and non-linearities is crucial for shaping the way neural networks remember their inputs, and that studying the problem only from a linear algebra perspective may be misleading. 
\end{sloppypar}

\section{Discussion}

The problem of short-term memory in recurrent neural network has been investigated using a variety of models, from discrete-time networks \cite{jaeger2001,ganguli}, to continuous-time networks \cite{hermansSchrauwen10a,Buesingetal10} and spiking networks \cite{maass2002real,wallace}. Using various approaches from statistics to information theory and dynamical systems, existing literature has mainly focused on three major questions:
\begin{itemize}
\item How does memory relate to the number of nodes in the network ?

Since \cite{jaeger2001}, it has been shown that the relationship between the memory capacity and the number $n$ of nodes in the network is essentially linear. Beyond this linear relationship between the memory span and $n$, a recent study \cite{charles} has shown the ability of linear recurrent network to perform a compressed sensing operation and to achieve exponentially long memory for sparse inputs, echoing ideas introduced in \cite{ganguli2010short}.

\item What is the role of non-linearities ?

While short-term memory in linear random recurrent networks has been studied extensively in \cite{ganguli,HermansSchrauwen09}, the case of non-linear models is less well understood as the increase of the amount of recurrence (e.g. through the special radius of the connectivity matrix) controls simultaneously the memory and the amount of nonlinearity in the representation. This trade-off between memory and nonlinearity has been investigated in particular from a theoretical perspective in \cite{Dambreetal12}, which shows how these two components can be defined and measured, and how it imposes constraints on the overall performance of reservoir computing systems.

\item What is the impact of the connectivity structure ?

As discussed in Section 2, the impact of connectivity structure on memory has been investigated in \cite{ganguli}, showing the importance of non-normality of $\W$. In \cite{White,HermansSchrauwen09}, the specific case of orthogonal connectivity matrices is also studied, showing the robustness of such structures to noisy perturbations, a type of matrix also under consideration in \cite{charles} to demonstrate the RIP property. Furthermore, in a series of articles \cite{ozturk2007analysis,song2010effects,rodan2011minimum,Straussetal12}, several authors have explored the impact of connectivity structure in terms of prediction performance, showing that simple deterministic connectivity, such as linear chains, may perform very well in various tasks. The relationship between memory and performance is not straightforward as it may be very task-dependent. However, a recent study \cite{couillet} of the performance of linear ESN has identified a connection between the Fisher memory curve of \cite{ganguli} and the mean-square-error prediction performance. Finally, various works have been interested in understanding the interplay between connectivity structure and autonomous non-linear reservoir dynamics (e.g. \cite{garcia,wainrib2015regular}) but not in the perspective of understanding short-term memory properties.
\end{itemize}

The interplay between connectivity structure and non-linearity may have important consequences for short-term memory and theoretical studies of this problem remain scarce. The present theoretical analysis of short-term memory and context-dependent representation in recurrent neural networks, although limited by its modeling assumptions (choice of the dynamical system, connectivity models), has contributed to advance the understanding of this phenomenon: 
\begin{enumerate}
\item Since \cite{ganguli}, it is known that {in linear models, the distinction between normal and non-normal connectivity matrices is very important}. 

\item Similar results hold for other memory measures (mutual information and context capacity) in the linear case: random {symmetric connectivities capture less information} about the input.

\item However, we have shown that this is no more the case in non-linear models : {in the non-linear regime, symmetric random RNN outperform the asymmetric model}. 

\item Mean-field theory is reaching its limitation: we have shown how it can be used for the asymmetric model, but it does not provide the key to unlock the symmetric one (more generally, structured models).
\end{enumerate}

\section*{References}

\bibliographystyle{apalike}
\bibliography{biblio}
\end{document}